\documentclass[twocolumn,showpacs,preprintnumbers]{revtex4}
\usepackage{amsfonts}
\usepackage{amsmath}
\usepackage{graphicx}
\usepackage{dcolumn}
\usepackage{bm}
\usepackage{amssymb}

\begin{document}
\title{Dynamics of Charged  Events}
\author{Constantin Bachas${}^{\, \sharp}$, Claudio Bunster${}^{\, \flat}$ and Marc Henneaux${}^{\, \natural\, \flat}$\vspace{5mm}}
\affiliation{${}^{\sharp}$ Laboratoire de Physique Th\'eorique de l'Ecole Normale Sup\'erieure
 {\rm \cite{afiliation}} 
 \\ 24 rue Lhomond, 75231 Paris cedex, France}   \vskip 2mm
\affiliation{${}^{\, \natural}$ Physique Th\'eorique et Math\'ematique, Universit\'e Libre
de Bruxelles and International Solvay Institutes,\\ ULB Campus Plaine C.P. 231,  1050 Bruxelles, Belgium\\  
${}^{\flat}$ Centro de Estudios Cient\'{\i}ficos (CECS)\\  Av. \hskip -2mm  Arturo Prat
514, Valdivia, Chile.}  
 
\date{\today }

\begin{abstract}
In three spacetime dimensions the worldvolume of a magnetic source is a
single point, an event. We make the event dynamical by regarding it as the
imprint of a flux-carrying particle impinging from an extra dimension. This can be
generalized  to higher spacetime dimensions and to extended events. We exhibit
universal observable consequences of the existence of events and argue that
events are as important as particles or branes. We explain  how events arise
on the worldvolume of membranes in M theory,  and in a Josephson junction
in superconductivity.
\end{abstract}

\maketitle
The principle of electric-magnetic duality states that electric and magnetic
fields must be treated on equal footing. It implies that one must consider both
electric and magnetic sources in the Maxwell equations.

When the dimension of the electric source is increased, the dimension of the dual
magnetic source  decreases. In the extreme case of an electric
object whose world-volume has dimension D-2, in D spacetime dimensions, the
worldvolume of the corresponding magnetic object has dimension
zero.  The worldvolume is just a point.  We shall say then that  the magnetic
source is an \textit{event}.  The simplest example of  an event is the dual of
an ordinary point electric charge in D=3  spacetime dimensions.  
Endowing 
such an object with dynamical properties seems to be a contradiction in terms.
Thus, so far,  events have been considered as external sources \cite{CT1,Henneaux:1986tt}. 
This
is unsatisfactory because it violates  the principle of action and reaction:
the source acts on the electromagnetic  field, but the field does not react on the source. 
In order to have a
closed physical system one
must,  therefore,   assign dynamics to the source.
 A way to achieve this is the following:  One considers the
event  as the imprint of  a  flux-carrying particle  which  comes  in from an extra
dimension,  hits spacetime and spreads its flux over it.  It will be shown below
that this idea can indeed be carried through successfully.

  The idea uses two key ingredients:  (i)  The 
  electromagnetic  field must be localized on a  D=3 slice of  a higher-dimensional spacetime, and
  in addition (ii) magnetic flux can  escape in the extra dimensions  where it is carried 
    by point-like particles.   
  Two examples  in which these ingredients are present  include the membrane of M theory and
  S-I-S  Josephson junctions.  We will  identify dynamical events in these familiar systems.

Solving in this way the problem of the dynamics of a point source in
spacetime  opens a treasure chest. A point event is the 
simplest member of a family which includes also
\textit{extended events}. An extended event is the imprint left on spacetime
when a p-brane with $p\geq 1$  hits it.    Just as point events,
extended events do not obey an equation of motion in the D-dimensional spacetime,
 because their
dynamics takes place in the extra dimensions.    This is the key difference
between events and particles,  or  ordinary extended objects. For example a
closed one-dimensional event is a loop in the D-dimensional spacetime with arbitrary  shape,
it can be timelike, spacelike, null \cite{bh} or a combination of all
three. It can also go forward and backwards in time with no violation of
causality. The  worldline is not a history. However it still produces fields in 
the D-dimensional spacetime and
interacts with them within  a closed physical system.

It is the main point of this article to put forward the idea that events
should play as central a role\ in Physics as that played by particles or
ordinary branes.  It should be stressed here that events occur in Lorentzian spacetime
and  should be distinguished from instantons, i.e.  solutions of the  Euclidean equations
that describe tunneling.  {This  will become more clear at the end.}

{} With  no essential loss of generality we will  restrict ourselves  to
the magnetic point events of  D=3  electrodynamics, which motivated
the introduction of the concept. The 
generalization to extended
events will be  straightforward. {A simple }  action is the sum of three terms,
\begin{eqnarray}\label{action}
I =  & -\frac{1}{2}\int d^{3}x\ ^{\ast} \hskip -0.7 mm F_{\mu}
\hskip -1mm \ ^{\ast}\hskip -0.7 mm F^{\mu} -  m\int^{0} d\tau  \sqrt{-\dot{z}_{M}\dot{z}^{M}} 
\cr &\  \cr    &+\,  g\int d^{3}x\, \delta^{(3)}(x-\bar z)\varphi(x)\ . 
\end{eqnarray}
Here  $\hskip -0.7mm \ ^{\ast}F_{\mu}=\partial_{\mu
}\varphi$ with $\varphi$ the magnetic potential, and $z^M(\tau)$ is the worldline of the flux-carrying particle
coming in from the past
  and absorbed at $\tau= 0$ . The coordinates of the event are, in self-explanatory notation, 
   $z^M(0) = (\bar z^\mu, 0)$.
   The third axis runs along the extra dimension and the signature is $(-,+,+,+)$.
   For a particle
being emitted {(an inverse event)}  
one should reverse the  limits of the $\tau
\ $integration  as well as  the sign of $g$.

{} The equation of motion for $\varphi$ reads 
\begin{equation} \label{field}
\square\varphi=g\delta^{(3)}(x-\bar z)\ . 
\end{equation}
The general solution of\ (\ref{field}) is the sum  $\varphi
=\varphi_{0}+gG$, where $\varphi_{0}$ is the general solution of the
homogeneous equation $(g=0)$ and $G$ is a Green function of the wave operator.
At the classical level a natural choice is to take for $G$ the retarded Green
function which vanishes outside the future  light cone of the event,%
\begin{equation}
G_{R}(x\text{)=}\frac{1}{2\pi}\frac{1}{\sqrt{-x_{\mu}x^{\mu}}}\theta
(- \mathrm{x}_{\mu}\mathrm{x}^{\mu})\theta(\mathrm{x}^{0})  \label{Green}%
\end{equation}
where we have here set  $\bar z = 0$.  If one takes $\varphi_0 = 0$ 
the situation as seen
from within the  D=3 spacetime  is the following:
For $x^{0}<{0}$ there is no field.
\ At $x^0 = 0$, suddenly a flash of   light emerges and propagates to
the future \cite{flash}. 
This situation will be the most probable classically because it corresponds
to  a flux-carrying particle impinging from
the extra dimension, without any precise control  
 (\textquotedblleft
fine tuning \textquotedblright) 
 of its initial conditions \cite{Durin}.   The time-reversed process, 
where one would replace the retarded Green function
$G_{R}$ by the advanced one $G_{A},$ corresponds  to  a precisely prepared (\textquotedblleft
fine tuned\textquotedblright) pulse of radiation converging on the spacetime point $\bar z= 0$, 
and disappearing as  a flux-carrying particle into the extra dimension.
This situation would be classically improbable. 

\vskip 1mm 
The action \eqref{action}  is not fully satisfactory because 
it does not allow a proper treatment of energy-momentum balance.   
Firstly,   the energy of the event field  \eqref{field} diverges.
In order  to make  it finite one must smear the magnetic  source  over a finite region 
of the D=3 spacetime.
This smearing should be physically related to the  
approach and  absorption of  the flux-carrying particle at
 $z^3=0$.    There appears to be no natural  description
 of this process, consistent with Poincar\'e  invariance,  within the present simple effective model. 
 We therefore 
prefer to  leave the  issue of energy-momentum conservation
 to an eventual  underlying microscopic theory.  An analogous  difficulty actually arises
when one tries to compute  the mass of a  magnetic pole within the Dirac theory. 
 
A  more satisfactory situation is encountered in connection with the conservation
of  magnetic  flux and of angular momentum. The
 action \eqref{action}   is invariant, up to a surface term, under the global gauge
transformation $\varphi\rightarrow\varphi+c$, where $c$ is a constant.
The first two terms of   \eqref{action}   are obviously invariant, so one needs to analyze only
 the change in the third, coupling term. This  is just
$-g$ times the shift $c$.  In order to apply Noether's theorem one rewrites
this  change as  $c$ times the integral over the extended spacetime of the divergence of a current, 
i.e.  $ \Delta I =  c \int d^4x \,  \partial_{M}
J_{\rm particle}^{M}$ where
\begin{equation}
J_{\rm particle}^{M} = g\int_{-\infty}^{0} d\tau\,  \delta^{(4)}(x-z(\tau))\frac{dz^{M}}{d\tau
}\ . 
\end{equation}
It is easy to verify that  $\partial_{M}
J_{\rm particle}^{M}=-g\delta^{(4)}(x- \bar z),$
and  $\int d^{3}x\ J_{\rm particle}^{0}= g\,  \theta (\bar z^0 - x^0)$ with $\theta$ the step function.
Furthermore, the contribution to the current from the Maxwell term is given by 
\begin{equation}
J_{\text{field}}^{M}=  (\hskip -0.5mm \ ^{\ast}F^{\mu},0) \ \delta(x^{3})\ . 
\end{equation}
On account of the field equation (\ref{field}) this obeys 
$\partial_{M}J_{\rm field}
^{M}=g\delta^{(4)}(x-\bar z)$ and 
$\int d^{3}x\ J_{\rm field}^{0}=  g\,  \theta (x^0 - \bar z^0)$. The divergences of the two currents cancel, 
and one  obtains the  flux
conservation law  $\partial_{M}(J_{\text{particle}}^{M}+J_{\text{field}}^{M})=0$.

In conclusion,  the conserved charge due to the (global)  gauge invariance
of the action is the magnetic flux $g$, which is being  transferred to the electromagnetic field
 by the impinging particle. 
 Unlike the situation with energy and momentum, this deposited  flux is
finite. It  constitutes the main signature of the event, independent of any  microscopic details.
 
The most striking consequence of the existence of  magnetic events is the
quantization rule for the product of the electric charge $e$ and of the event
charge $g$,
\begin{equation}
eg=2\pi\hslash n,\ \ \ \ \ \ \ \ \ \ \ \ \ \ \ \ n\mathrm{\ }\text{integer}\ . %
\label{1.1}%
\end{equation}
{}For the magnetic pole in 3+1 spacetime dimensions, this result was first
obtained by Dirac in 1931, and rederived by him  in 1948 from the quantum
mechanical implications of a classical point-particle action principle. Later
on Dirac's 1948 derivation was proven to remain valid for extended objects
\cite{CT1}. The quantization rule (\ref{1.1}) has survived the embedding of
the effective point-particle theory in detailed microscopic models. Since its
derivation is independent of whether the electric and magnetic
sources obey equations of motion or not, it remains valid for events of any dimension.

 The fact that magnetic events imply the quantization of electric charge is,   conceptually, even
 more striking than the analogous implication of magnetic poles.  Indeed,  for
example in four-dimensional space time, one could have in the whole history of
the universe a single magnetic event, which could be a very small loop
carrying a quantum $g$ of magnetic flux. That small loop could lie entirely in
the \ very far past (or, for that matter, in the very far future) and yet all
electric charges, at all times, should have to be quantized according to
(\ref{1.1}). One could say that  magnetic events are like  the Cheshire cat,  they vanish leaving only
their smile behind.



There exists  one more property of magnetic events which, like the conservation of magnetic
flux and the quantization rule of electric charge, is simple and universal, i.e. it is 
 expected to 
survive in any underlying microscopic theory. This is angular momentum. 
Consider an electric particle with worldline $y^\mu(\tau)$. The particle starts feeling
the influence of the event after it enters its future light cone. The conserved angular
momentum  is the sum of the orbital angular momentum of the
electric particle,  $L^{\mu\nu}$, and of the angular momentum stored in the electromagnetic  field,
$J^{\mu\nu}_{\rm field} =  \int_{\Sigma
}\,(T_{\mu}^{\ \nu}\,x^{\rho}-T_{\mu}^{\ \rho}\,x^{\nu})\,d\Sigma^{\mu}\ .$  The latter
vanishes before the interaction, and it is given afterwards  by 
\begin{equation}
J^{\mu\nu}_{\rm field}  = 
eg/2\pi\ \left(  y_{\rho}%
/\sqrt{-y^{2}}\right)  \varepsilon^{\mu\nu\rho}\ . 
\end{equation}
The electromagnetic field whose energy-momentum tensor enters in the expression for
$J^{\mu\nu}_{\rm field}  $ is the sum of the retarded field \eqref{Green} of the magnetic event, 
 and of the field of the
electric particle. The only contribution comes from the cross term. 
To evaluate the integral it is necessary to regularize the event field on the light-cone.
The result uses only the Gauss law and does not depend on the details of the
regularization. 

 Since total angular momentum is conserved, the change in the orbital angular
 momentum of the particle is given by $\Delta L^{\mu\nu} = - J^{\mu\nu}_{\rm field}$
 \cite{CBW.Mart}. 
The dual pseudovector  $^{\ast}\Delta L^{\mu}=\frac{1}{2}\epsilon^{\mu\nu\rho}\Delta L_{\nu\rho}$
 points from the event to the electric charge and, in view of the quantization rule \eqref{1.1},
 its magnitude is   an {\it integer} multiple of $\hbar$, 
 \begin{equation}\label{angm}
 \vert  \hskip -0.9mm \ ^{\ast}\Delta L \vert =  
eg/2\pi\  = n \hbar \ . 
\end{equation}
This result mirrors the well-known fact that, in 3+1 dimensions, the angular momentum
stored  in the field
of an electric and a magnetic charge is a multiple of $\hbar/2$. 
 \vskip 1mm
 
 As the last development in the context of the effective classical theory \eqref{action} of events,   
 one may consider the addition of a
 Chern-Simons term to the Maxwell action. The modified field equations take the form
\begin{equation}\label{15}
\partial_\mu F^{\mu\nu} +  \kappa   \hskip -0.7mm \ ^*F_\nu = j^\nu_{\rm el}
\ \   \ \ \ {\rm and}\ \ \ \ \  \partial_\mu\hskip -1.2mm \ ^*F^{\mu} =  j_{\rm mag} \ ,
\end{equation} 
where $\kappa$ is the coefficient of the Chern-Simons term and we have here allowed for
generic sources.  Taking the divergence of the left-hand equation gives
$  \partial_\nu j^\nu_{\rm el} \, = \,  \kappa \,  j_{\rm mag}$.  
Hence, a magnetic event   deposits automatically   $\kappa g$ units of electric charge.   
Furthermore, since this latter is  quantized, one concludes that $\kappa$ must
 be an integer multiple
of  $ {2\pi \hbar/ g^2}$. It is worthwhile noting
here   that the presence of magnetic events  in 
a Chern-Simons theory not only quantizes
electric charge, \textit{but requires its existence}!

Although, in 2+1 dimensions the quantization of the Chern-Simons term is
automatic if the photon field comes from a spontaneously-broken non-abelian
gauge theory \cite{mass}, this property was first established, in terms of
what we now call an event,\ in \cite{Henneaux:1986tt}. 
For higher dimensions
the quantization of the coupling was derived in \cite{Bachas:1998rg} (see also
\cite{BG}) by using dimensional reduction and the Witten effect, namely  the
acquisition of electric charge in the presence of a topological $\theta$-term.
One   arrives  at the same conclusion without having to dimensionally reduce
the theory, along lines that generalize the analysis in $2+1$ dimensions.
The argument in D$=2n+1$ dimensions uses $n$ extended events whose
codimensions span the entire space.
  
So far  we have dealt with universal properties of events as described by 
 a simplified effective theory.  Next we will exhibit two examples of their ocurrence
in instances where the underlying microscopic theory already exists.
 

The first example  is the impinging of D-particles on D2-branes in string theory.
The Mawxell field lives on the worldvolume of the D2-brane, which is in turn embedded
in a ten-dimensional spacetime  \cite{Pol}. Bulk D-particles feel 
 the gravitational attraction of the D2-brane, but this is negligibly weak at
distances $\gg{\ell}_{s}$, the characteristic string length. 
 However, whenever a D-particle gets sufficiently
close to the D2-brane,   a tachyonic instability develops
\cite{Banks:1995ch} and it gets rapidly absorbed. 
  To an observer on the
 D2-brane the process looks precisely like a magnetic event: the particle
deposits a flux $\int F=T_{0}/T_{2}$, which spreads
at the speed of light. Here $T_{0}$ is the  mass of the D-particle
 and $T_2$ is the  tension of the  D2-brane.
  If one
normalizes canonically the photon field  the magnetic charge of the event is
$g=T_{0}/\sqrt{T_{2}}\ $. The
 minimal electric charge, which corresponds to
the endpoint of an open string, is given by $e=T_{F}/\sqrt{T_{2}}$, where
$T_{F}\equiv\hbar/2\pi\ell_{s}^{2}$ is the fundamental-string tension. 
 Using Polchinski's values  for the
 tensions  \cite{Pol} one then finds, $eg\,=\, 2\pi\hbar\ .$ 
The capture of a D-particle
corresponds, indeed, to a magnetic event with the minimal allowed charge.
 
 One may view the impinging D-particle as a small spherical D2-brane carrying
 one unit of magnetic flux. Events and inverse events acquire then a geometric
meaning, as the joining and splitting of membranes.  It is furthermore possible to induce a
 Chern-Simons coupling  by inserting one (or more) 
 D8-branes between the D2-brane and the Minkowski vacuum   \cite{Green}.   When the D-particle
 traverses a  D8-brane an open  string stretching between them is automatically 
  created \cite{Bachas:1997ui}.  
After the D-particle
has been dissolved, these open strings remain attached to the D2-brane. They are the
electric charges deposited by the event. All universal features of events
are, thus,  neatly illustrated in this string theory example.   
 
\vskip 0.8mm


 Our second example is taken from the theory of superconductivity. The setup
 is that of a S-I-S Josephson junction, consisting of a thin
insulating layer sandwitched between two superconducting bulk pieces.
The (history of the)  insulating layer will be our D=3 spacetime, while the
adjacent (history of the) superconductor will be the extended spacetime.
This setup was proposed in  \cite{Tetradis:1999tn} as an example of 
``localization"  of the Maxwell field. To see why, recall that the superconductor
may be described to a first aproximation as a macroscopic quantum
fluid of Cooper pairs of charge $q=2e$, where $e$ is the
charge of the electron.  The Cooper pairs condense in the ground state, 
as   expressed by the non-vanishing expectation value of a complex scalar field, 
 $\langle \Psi \rangle = \vert \Psi_0 \vert e^{i\omega}$. This condensate screens 
all  electric charges and  expels the magnetic field  from the bulk. The 
longitudinal electric field and the normal magnetic field continue to vanish
inside the  insulating layer, while the remaining non-zero fields  
can be identified as those of the effective D=3 electrodynamics, 
 \begin{equation}\label{ide}
 \ ^{\ast}F^\mu\,  \equiv\,  {\rm F}^{3\mu} \, =\,  \frac{1}{2}\epsilon
^{\mu\nu\rho}\ ^{\ast}{\rm F}_{\nu\rho\,}.
\end{equation}
Here math-style  and text-style symbols denote the three- and four-dimensional
fields. The latter are evaluated at the position ($x^3=0$)  of the junction,  
which is sufficiently thin so as to neglect, in its interior, all  field variations
along the $x^3$ direction.
Notice that the identification  (\ref{ide})  has exchanged   the roles 
of ``electric"  and ``magnetic".
      
\eject
      The ``magnetic"  events of the D=3 theory correspond to electric
charges  crossing  the insulating layer. This follows from  (\ref{ide})  and   the Maxwell
equation $\partial_\mu {\rm F}^{3\mu} =  - {\rm J}^3 $.  Integrating over the junction
gives, for an isolated system,  the conserved global charge 
 $\ \int d^2x \hskip -0.5mm  \ ^{\ast}F^0   + Q_{\rm right} $, where $Q_{\rm right}$
is the charge in the right-hand side of the junction. 
 The ``magnetic" flux,  $g$,  
deposited by the event is therefore equal to the total electric charge transferred. The effective D=3 theory
has also  ``electric" point-like charges  if the superconductors are of type II.
These  charges are the endpoints of Abrikosov vortex tubes which spread their
magnetic flux in the insulating  layer.  The quantum of flux, $\Phi_0 = h/q$,  is the effective
``electric''  charge in three dimensions. The quantization condition \eqref{1.1}  is thus valid, 
provided  the charge  carriers crossing the junction are Cooper pairs \cite{comment}.

    The most striking feature of a Josephson junction is the existence of a  (Josephson)   
  current  at zero voltage drop \cite{Tinkham}. This 
  arises because of the quantum tunneling  of Cooper pairs across the insulating layer.
  The effect is described mathematically by  {\it instantons}, which are the solutions  of
  equation \eqref{field} in imaginary time.  The action of an instanton controls the
  strength of the Josephson current, ${\rm J}^3 = {\rm J}_{\max}\sin ({q\varphi/ \hslash})\ $ with
  ${\rm J}_{\max} \sim e^{-I_{\rm inst}}$, as well as the non-perturbatively-small  mass of $\varphi$
  in the effective three-dimensional theory.

  Although closely-related,  instantons should be distinguished from  real-time events. 
  Which of the two is more relevant depends on the detailed physical context. The example of
  the D2-brane in  string theory helps to  illustrate clearly this point:  an event describes the 
  capture of a D-particle, whereas  instantons  describe the tunneling of flux from a neighboring
  D-brane.  Thus, an observer living on a D2-brane in a  D-particle gas will witness 
  numerous real-time events,  but there will be no tunnelings if  there is no other 
  D-brane nearby. 
   
 We conclude by remarking that, even though superconductivity makes events tangible,
 the more striking implications of their existence  might reside in astrophysics and cosmology.  
 The ideas of this letter cannot, however,  be simply  transposed to gravity. 
 Nevertheless,  the distinction between instantons and events acquires 
  special meaning  in cosmology:  Was our Universe created out of nothing,  by quantum
  tunneling, or was the Big-Bang an extended (and most dramatic)  event?
   \cite{Khoury:2001wf}

\textit{Acknowledgments.-- }We would like to  thank Roland Combescot, Denis Bernard
 and Jan Troost for useful discussions. C. Bunster
thanks Vivian Scharager for her essential assistance in the preparation of the
manuscript. C.~Bunster and M.~Henneaux are grateful to the 
Laboratoire de Physique Th\'eorique de l'Ecole Normale Sup\'erieure  for kind
hospitality. Work supported in part by IISN-Belgium, by a FP6 Network grant of
the European Union, and by the Belgian Federal Science Policy Office. The
Centro de Estudios Cient\'{\i}ficos (CECS) is funded by the Chilean Government
through the Millennium Science Initiative and the Centers of Excellence Base
Financing Program of Conicyt.


\begin{thebibliography}{99}
\bibitem{flash}  In the example of the Josephson junction, discussed at the end of the paper, this flash
of (linearly polarized) light could be conceivably observed.
\end{thebibliography}
\end{document}